# Small-signal amplifier based on single-layer MoS$_2$

*Branimir Radisavljevic, Michael B. Whitwick and Andras Kis*[*]

Electrical Engineering Institute, Ecole Polytechnique Federale de Lausanne (EPFL), CH-1015 Lausanne, Switzerland

*Correspondence should be addressed to: Andras Kis, andras.kis@epfl.ch

ABSTRACT. In this Letter we demonstrate the operation of an analog small-signal amplifier based on single-layer MoS$_2$, a semiconducting analogue of graphene. Our device consists of two transistors integrated on the same piece of single-layer MoS$_2$. The high intrinsic band gap of 1.8 eV allows MoS$_2$-based amplifiers to operate with a room temperature gain of 4. The amplifier operation is demonstrated for the frequencies of input signal up to 2 kHz preserving the gain higher than 1. Our work shows that MoS$_2$ can effectively amplify signals and that it could be used for advanced analog circuits based on two-dimensional materials.

MANUSCRIPT TEXT

The limits of further miniaturization of silicon nanoelectronic devices lead to an emerging development of novel nanomaterials. It is anticipated that silicon complementary metal-oxide semiconductor (CMOS) technology will reach the scaling limit in the near future. Electronic materials, such as semiconductor nanowires,[1] carbon nanotubes,[2] III-V compound semiconductors[3] are being proposed as promising candidates to replace silicon in electronic devices. On the other side, two-dimensional materials, such as graphene and transition-metal chalcogenides, are attractive for novel nanoelectronic devices because, compared to working with one-dimensional materials it is relatively easy to tailor them in desired shape and fabricate very complex structures from them. Being just few angstroms thick, 2D materials are very attractive for new generation transistors. Scaling theory predicts that short-channel effects, such as threshold voltage roll-off, drain-induced barrier lowering and impaired drain current saturation, in field-effect transistors should be much less pronounced in this case.[4] Graphene,[5] an atomically thin two-dimensional sheet of graphite is very promising because of its extraordinary properties, such as high carrier mobility up to 230 000 $cm^2/Vs$,[6] and two-dimensional geometry that is at the ultimate limit in vertical device scaling. In its pristine form, graphene does not have a bandgap that is of crucial importance for realization of field-effect transistors (FETs) with satisfactory on/off ratios. We have recently shown[7] that field-effect transistors based on monolayer $MoS_2$ have room temperature mobility comparable to that of the best graphene nanoribbons fabricated to date[8] and strained thin silicon films,[9] with current on/off ratio higher than $10^8$.

Single-layer $MoS_2$, 6.5 angstroms thick, is a 2D direct gap semiconductor that can be exfoliated from bulk crystal using scotch tape micromechanical cleavage,[10] lithium intercalation[11] or liquid exfoliation methods.[12] Its energy bandgap of 1.8 eV[13] makes it very suitable for nanoelectronic applications, such as field-effect transistors,[7] and devices based on them, such as digital logic gates[14] and analog small-signal amplifiers. Single-layer MoS$_2$ is also 30 times stronger than steel[15] which makes is suitable for use in flexible electronics. Additionally, by decreasing the number of layers in MoS$_2$ crystal stack down to monolayer, an indirect bandgap of 1.2 eV converts to a direct one of 1.8 eV, giving an opportunity for engineering new optoelectronic behaviors and gives promise for new nanophotonic applications.[16]

Here, we demonstrate the first realization of integrated voltage amplifier based on 2D semiconducting material MoS$_2$ with voltage gain *G* higher than 1, making it suitable for incorporation in analog circuits where amplification of small AC signals is necessary. Thanks to its very high input impedances, such amplifiers may be essential when dealing with the high-impedance signal sources in new nanoelectronic devices.

We start the fabrication of our transistors with scotch-tape micromechanical exfoliation[5] of single-layer MoS$_2$ on top of the degenerately doped silicon substrate covered with 270 nm thick SiO$_2$, as shown in Figure 1a. This oxide thickness has been shown to be the optimal one for optical detection of single-layer MoS$_2$.[17] The resulting single-layer MoS$_2$ is highly crystalline.[18] Electrical leads for the source and drain are fabricated using electron-beam lithography, followed by evaporation of 90 nm thick Au layer and standard metal lift-off procedure in acetone. In order to decrease contact resistance and remove resist residue devices are annealed in Ar/H$_2$ atmosphere at 200 ºC for 2 hours.[19] After that, devices are covered with 30 nm thick layer of HfO$_2$ deposited by atomic layer deposition (ALD). ALD is performed in a Beneq system using a reaction of H$_2$O with tetrakis(ethyl-methylamido)hafnium at 200 ºC. HfO$_2$ has a dielectric constant of ~19 and acts as a top-gate dielectric. Finally, another electron-beam step is performed for top gate electrodes, and followed by Cr/Au (10/50 nm) metallization. The typical structure of our devices, in this particular case composed of two transistors connected in series, is shown in Figure 1b. Cross-sectional view of a field-effect transistor based on single-layer MoS$_2$ is shown on Figure 1c.

Before connecting them in the amplifier circuit, we perform basic electrical characterization of our MoS$_2$ transistors. All electrical measurements are carried out using National Instruments DAQ cards, a home-built shielded probe station, and an Agilent E5270B parameter analyzer. In Figure 2 are shown electrical characteristics of the one of our transistors. The second transistor used for the realization of an integrated amplifier presented in this Letter has very similar electrical characteristics, a prerequisite for successful operation in an electrical circuit. All measurements are performed in the air at room temperature. First, we applied drain-source bias $V_{ds}$ to a pair of gold leads and back-gate voltage $V_{bg}$ to the silicon substrate which acts as a back-gate electrode, as it is highly p-doped. These gating characteristics are presented in Figure 2a, showing a typical behavior of FETs with n-type channel. Using the expression $\mu = [dI_{ds}/dV_{bg}] \times [L/(WCV_{ds})]$, where $L$ = 1.6 µm is the channel length, $W$ = 4.2 µm is the channel width and $C$ = 1.3 × 10$^{-4}$ F m$^{-2}$ is the back-gate capacitance per unit area ($C = \frac{\varepsilon_0 \varepsilon_r}{d}$; $\varepsilon_0 = 8.85 \times 10^{-12} \frac{F}{m}$; $\varepsilon_r = 3.9; d = 270$ nm), we extracted filed-effect mobility of ~380 cm$^2$/Vs. This is still a lower limit of mobility as it is a two-contact measurement and contact resistance is not excluded. Linear dependence of drain-source current $I_{ds}$ on bias voltage $V_{ds}$ clearly indicates the ohmic character of our gold contacts (Figure 2a, inset). The two-contact on-resistance is 25 kΩ for $V_{bg}$= 5 V and drain-source bias $V_{ds}$= 100 mV.

Local gate control of charge density in MoS$_2$ channel of our transistor is shown in Figure 2b. This is achieved via a Cr/Au top gate where the role of top-gate dielectric is played by a 30-nm thick HfO$_2$ ALD layer. During these measurements, the back-gate is kept grounded. For drain-source bias $V_{ds}$= 500 mV, we recorded a maximal on-current of 25 µA (5.94 µA µm$^{-1}$) and an off-current smaller than 5 pA (1.18pA µm$^{-1}$), resulting in an current on/off ration of ~ $2 \times 10^6$ in the ± 4 V top-gate voltage range. Subtreshold swing is ~500 mV/dec, reaching 150 mV/dec for some of our devices. In the inset of Figure 2b we present the dependence of drain-source current $I_{ds}$ on drain source bias $V_{ds}$ for different top-gate voltages, indicating efficient local-gate control of the channel resistance.

After basic characterization of the transistors we connect them in the amplifier circuit, as shown in Figure 3a. In this Letter, we demonstrate integrated common-source analog amplifier that basically has a function of amplifying small AC signal with negative gain $|G| > 1$. This circuit consists of two transistors connected in series, where one acts as a "switch" (lower one in Figure 3a) and the other one acts as an active "load" (upper one in Figure 3a). Gate of one of the transistors ("switch" transistor) acts as input, whereas gate of the other ("load" transistor) is connected with central lead and acts as output, Figure 3b. Power supply of the amplifier $V_{DD}$ is set to be 2V. The DC transfer characteristic of the amplifier based on the transistor from Figure 2 acting as a "load" is shown in Figure 3c. When a small AC signal $V_{in-AC}$ is superimposed on the DC bias $V_{gs}$ at the input, $V_{in} = V_{gs} + V_{in-AC}$, then under the right circumstances the transistor circuit can act as a linear amplifier. The transistor is first biased at a certain DC gate voltage to establish a desired current in the circuit, shown as the Q-point in Figure 3c. A small sinusoidal AC signal $V_{in-AC}$ of amplitude $\Delta V_{in}/2$ is then superimposed on the gate bias on the input, causing the output voltage $V_{out}$ to fluctuate synchronously with a phase difference of 180 degrees with respect to $V_{in-AC}$. The steepest region of the $V_{out}$ - $V_{gs}$ curve can be approximated by a straight line (red line in the Figure 3c) where the slope represents the voltage gain $G$ of the amplifier. In this case the gain of our amplifier is $|G| > 4$. It is important to note that for practical applications, a gain higher than 1 is desired.

To allow for maximum output voltage swing, the Q-point should be positioned approximately in the middle of the steepest region of the transfer characteristic in Figure 3c.

We achieve this by applying a DC bias of $V_{gs}$= 0 V to the top gate of our "switch" transistor and superposing AC signals $V_{in-AC}$ of different frequencies on this gate bias. We have applied small AC signal of 100 mV amplitude with frequency ranging from 30 to 2000 Hz shown in Figure 4a and 4b, respectively. It is important to notice that the AC signal is amplified and shifted in phase for 180 degrees, in agreement with the standard characteristic of common-source amplifiers. As shown in Figure 4c we performed measurements up to 2 kHz preserving the voltage gain higher than 1. For small frequencies (30 Hz) we can see that the gain $|G| = \Delta V_{out}/\Delta V_{in}$ is larger than 4. By increasing the frequency, the gain is reduced, reaching 1 at 2000 Hz. This is due to the influence of high parasitic capacitances and fringing fields that can be decreased by further circuit engineering, and by improving the device transconductance.

We have fabricated single-transistor amplifier as well, where the role of "load" plays a resistor of 30 MΩ (see Supplementary material[20]). The device consists of one transistor fabricated on monolayer $MoS_2$ flake connected in series with an off-chip load resistor. For some of our devices by using different load resistors and bias voltages $V_{DD}$ we were able to extract voltage gain $G$ larger than 10.

In conclusion, we have demonstrated an integrated small-signal analog amplifier based on single-layer $MoS_2$ filed-effect transistors. The amplifier is built using top gated $MoS_2$ transistors connected in series which with appropriate wiring forms an amplifier circuit with negative gain. Our amplifier exhibits a small-signal voltage gain higher than 4 and reaching 10 in the single-transistor configuration, surpassing graphene based two-dimensional amplifiers.[21] Operation of amplifiers with frequencies up to 2 kHz has been shown, with gain higher than 1 and no signal distortion. The frequency range can be further extended by decreasing parasitic capacitances of contacts, using exclusively on-chip wiring to connect the transistor gates and improving transistor transconductances. With a possibility of large scale production by solution-based processing or large-scale growth of $MoS_2$ thin films,[22] and the high mechanical strength of single-layer $MoS_2$,[15] our result could be very important for the realization of low cost and flexible small-signal amplifiers of new generation with higher integration densities.

ACKNOWLEDGMENT. Device fabrication was carried out in part in the EPFL Center for Micro/Nanotechnology (CMI). We thank T. Heine and G. Seifert for useful discussions, K. Lister (CMI) for technical help with the e-beam lithography system, and D. Bouvet (CMI) for support with ALD deposition. This work was financially supported by ERC grant no. 240076.

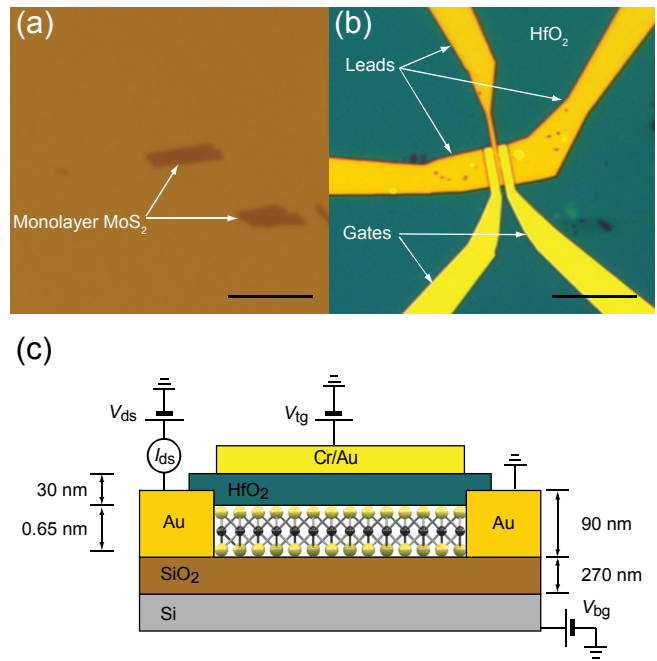

**Figure 1.** Fabrication of single-layer MoS$_2$ amplifiers. (a) Optical image of monolayer MoS$_2$ flakes deposited on top of a silicon chip with 270 nm thick SiO$_2$ layer. (b) Optical image of two field-effect transistors connected in series, fabricated on the upper flake shown in (a). (c) Cross-sectional view of a field-effect transistor based on single-layer MoS$_2$. Gold leads are used for the source and drain electrodes. The silicon substrate was used as a back-gate with the 270nm SiO$_2$ layer used as a dielectric. Top-gates were fabricated with Cr/Au leads and 30-nm thick HfO$_2$ dielectric. Scale bars in figure (a) and (b) are 10 µm.

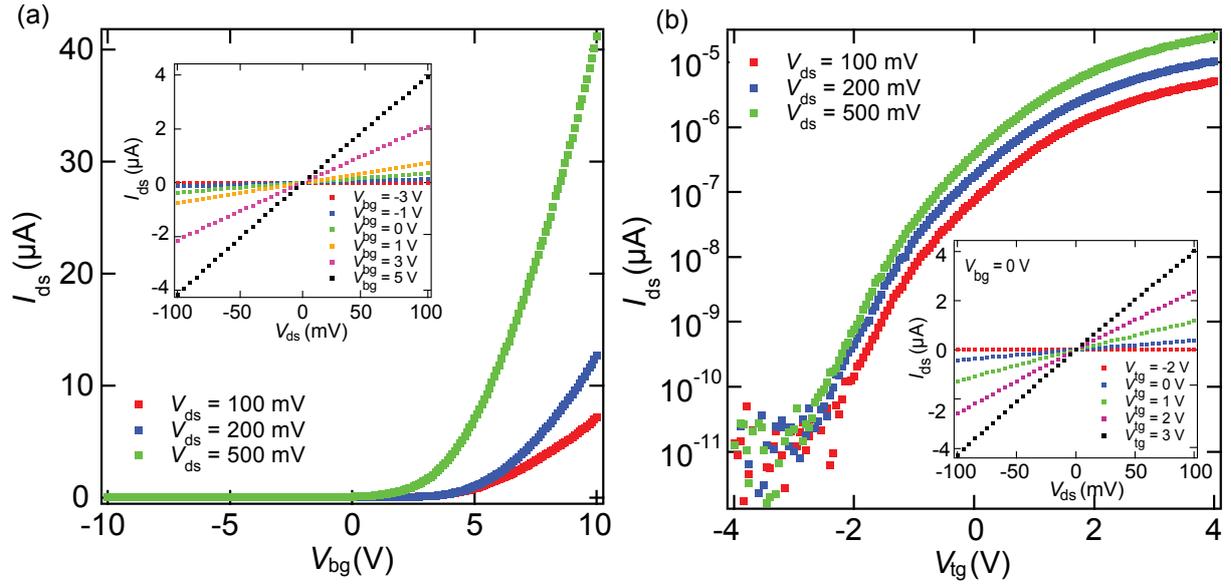

**Figure 2.** Electrical characterization of the field-effect transistors based on monolayer MoS$_2$. (a) $I_{ds}$-$V_{bg}$ curves acquired for different drain-source biases, indicating presence of n-type channel in our FET. Back-gate voltage is applied to the substrate and top-gate is disconnected. Extracted field-effect mobility from two-contact measurement is ~ 380 cm$^2$/Vs. Inset: $I_{ds}$-$V_{ds}$ curves for $V_{bg}$ ranging from -3 to 5 V. (b) Transfer characteristic of our FET for different source-drain biases. The device can be turned-off by changing the top-gate voltage from 2 to -2 V. The current on/off ratio is > 2×10$^6$ and subtreshold swing is ~ 500 mV/dec. Inset: $I_{ds}$-$V_{ds}$ curves recorded for different $V_{tg}$, with linear dependence clearly indicating ohmic gold contacts. Back-gate electrode is grounded.

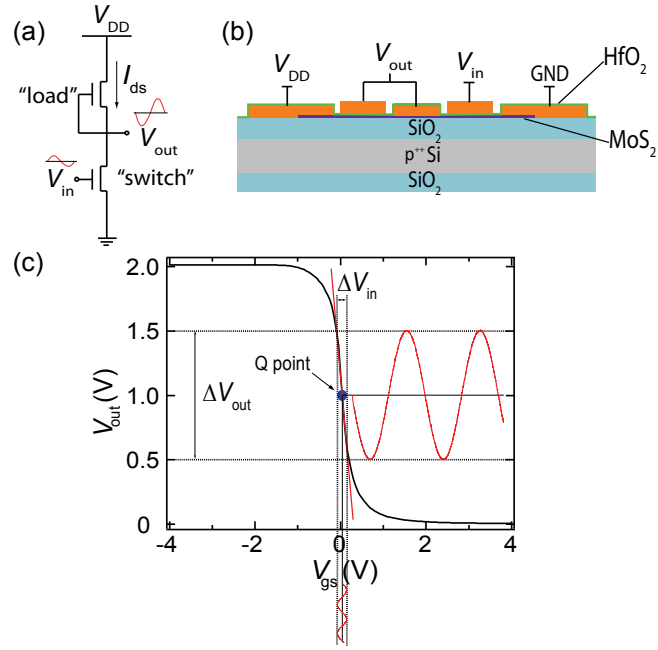

**Figure 3.** Transfer characteristic of the integrated MoS$_2$ amplifier. (a) Schematic drawing of integrated amplifier in common-source configuration. The lower transistor acts as a "switch" and upper as a "load". (b) Vertical cross-section of the amplifier device from Figure 1b with appropriate wire connections. (c) Transfer characteristic of the integrated amplifier realized with two transistors on the same MoS$_2$ flake. The "switch" transistor is first biased at a certain DC gate bias to establish a desired drain current, shown as the "Q"-point (quiescent point). At that point the "load" transistor has a certain constant resistance and acts as an active resistor. A small AC signal of amplitude $\Delta V_{in}/2$ is then superimposed on the gate bias of the "switch" transistor, causing the output voltage to oscillate synchronously with a phase difference of $\pi$.

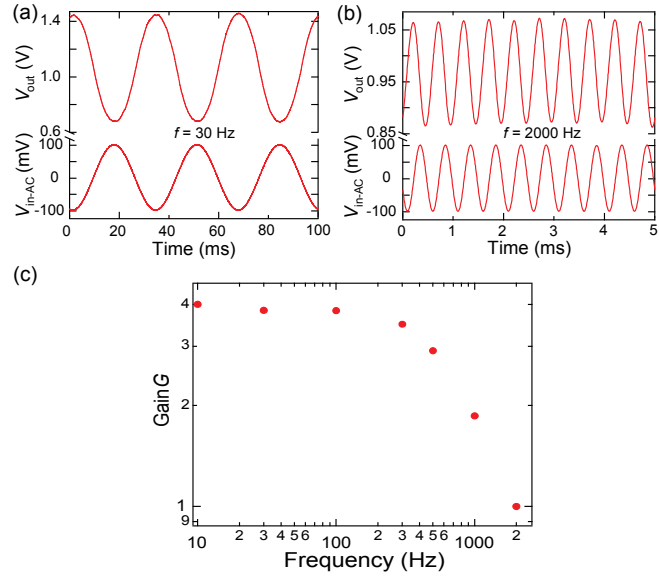

**Figure 4.** Demonstration of the small-signal amplifier operation. On the input of amplifier sinusoidal signal $V_{\text{in-AC}}$ of amplitude $\Delta V_{\text{in}}$= 100mV and frequencies of (a) 30 Hz and (b) 2000 Hz is applied with DC bias of $V_{\text{gs}}$ = 0 V, resulting in an amplified sinusoidal signal on the output. The output signal is shifted in phase for 180 degrees with respect to the input signal, in agreement with the standard characteristic of common-source amplifiers. (c) Voltage gain dependence on frequency of the input small signal. Voltage gain $|G| = \Delta V_{\text{out}}/\Delta V_{\text{in}}$ up to 100 Hz frequency of the input signal is ~ 4 and decreases with increasing frequency.

# Supplementary information

# Small-signal amplifier based on single-layer MoS$_2$

*Branimir Radisavljevic, Michael B. Whitwick and Andras Kis*[*]

Electrical Engineering Institute, Ecole Polytechnique Federale de Lausanne (EPFL), CH-1015 Lausanne, Switzerland

*Correspondence should be addressed to: Andras Kis, andras.kis@epfl.ch

Electrical characterization of the transistor that is used for realization of single-transistor MoS$_2$ amplifier is shown in Figure S1. Characteristics of the device are similar to the one shown for the transistor in the main text. The amplifier circuit consists of one transistor acting as a switch and one off-chip resistor R$_D$ acting as a load, connected in series, Figure S2a. Power supply of the amplifier $V_{DD}$ is set to be 3.5 V. The measurement procedure is the same as for the integrated amplifier in the main text.

The transistor is first biased at a certain DC gate voltage to establish a desired drain current, shown as the Q-point in Figure S2b. A small sinusoidal AC signal V$_{in-AC}$ of amplitude ΔV$_{in}$/2 is then superimposed on the gate bias $V_{gs}$, causing the drain-source current to oscillate synchronously. Fast slope of the $I_{ds}$ - $V_{gs}$ curve can be approximated by a straight line (red line in the Figure S2b) which slope represents transconductance of the amplifier, $g_m = \Delta I_{ds}/\Delta V_{in}$, that is in the case of our amplifier ~ 400 nS. The small-signal transconductance is crucial parameter for the gain of amplifier and high-frequency performance of the transistor. Knowing that the drain-source current $I_{ds}$ is related to the output voltage $V_{out}$ by $V_{out} = V_{DD} - I_{ds}R_D$, where $V_{DD}$ = 3.5 V is the amplifier supply, and $R_D$ = 30 MΩ is the load resistor, the amplitude of the AC output signal will be given by $\Delta V_{out} = -\Delta I_{ds} R_D = -g_m R_D \Delta V_{in}$. The voltage gain $G$ of the amplifier is therefore $G = -g_m R_D$, that is in the case of our amplifier $|G| > 10$, Figure S3a.

In Figure S3b is shown the possibility of tuning the gain of the amplifier by changing the power supply $V_{DD}$ while keeping the same load resistor $R_D$. We can notice that gain $G$ of the amplifier is increasing linearly with power supply $V_{DD}$.

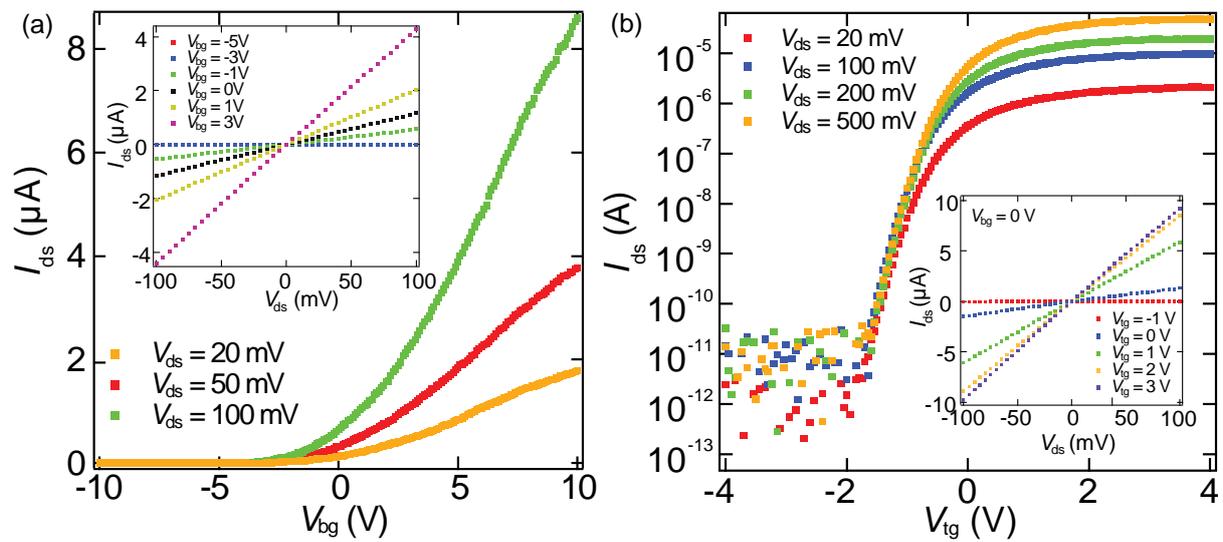

**Figure S1.** Electrical characterization of the field-effect transistor based on monolayer $MoS_2$. (a) $I_{ds}$-$V_{bg}$ curves acquired for different drain-source biases, indicating presence of n-type channel in our FET. Back-gate voltage is applied to the substrate and top-gate is disconnected. Inset: $I_{ds}$-$V_{ds}$ curves for $V_{bg}$ ranging from -5 to 3 V. (b) Transfer characteristic of our FET for different source-drain biases. The device can be turned-off by changing the top-gate voltage from 0 to -2 V. The current on/off ratio is $> 9 \times 10^6$ and subtreshold swing is ~ 200 mV/dec. Inset: $I_{ds}$-$V_{ds}$ curves recorded for different $V_{tg}$, with linear dependence clearly indicating ohmic gold contacts. Back-gate electrode is grounded.

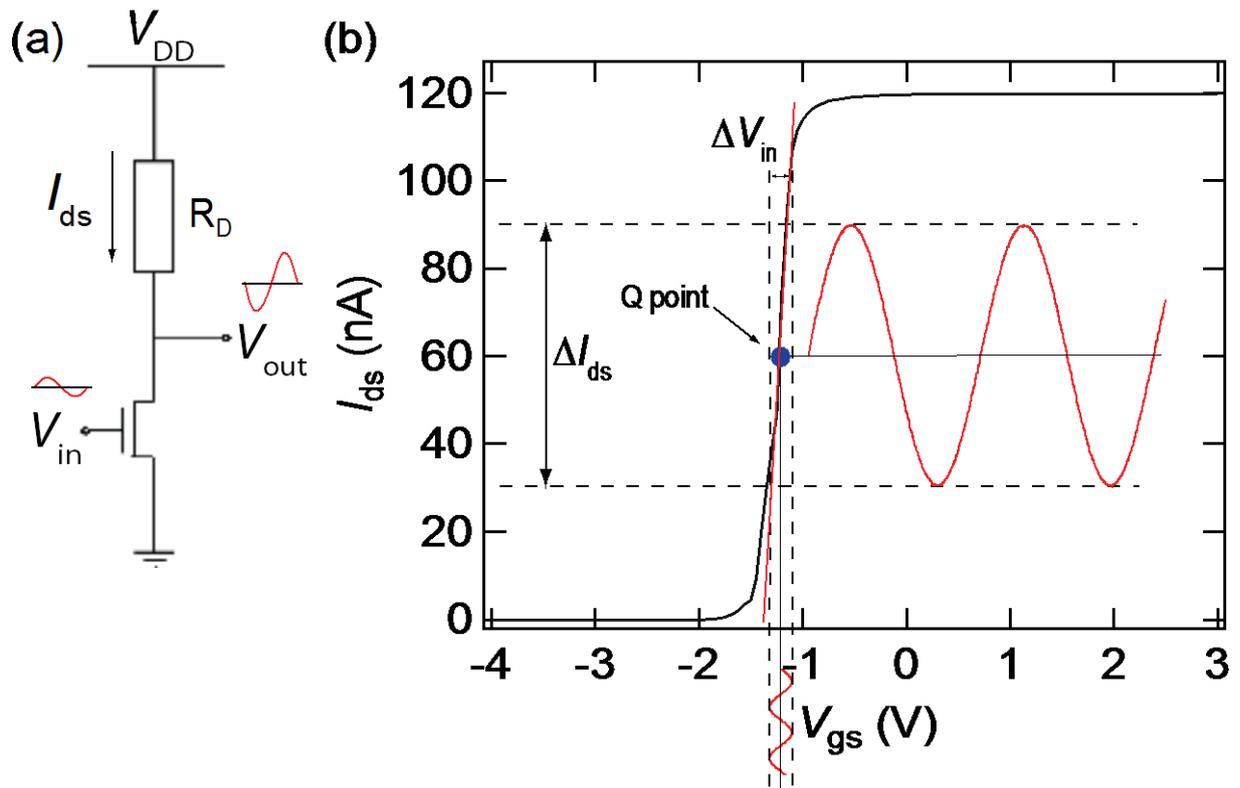

**Figure S2.** Transfer characteristic of the MoS$_2$ single-transistor amplifier. (a) Schematic drawing of single-transistor amplifier in common-source configuration. $R_D$ is an off-chip resistor chosen in this case to be 30 MΩ. (b) Transfer characteristic of the amplifier realized with transistor from Fig. 2. The transistor is first biased at a certain DC gate bias to establish a desired drain current, shown as the "Q"-point (quiescent point). A small AC signal of amplitude $\Delta V_{in}/2$ is then superimposed on the gate bias, causing the drain-source current to oscillate synchronously.

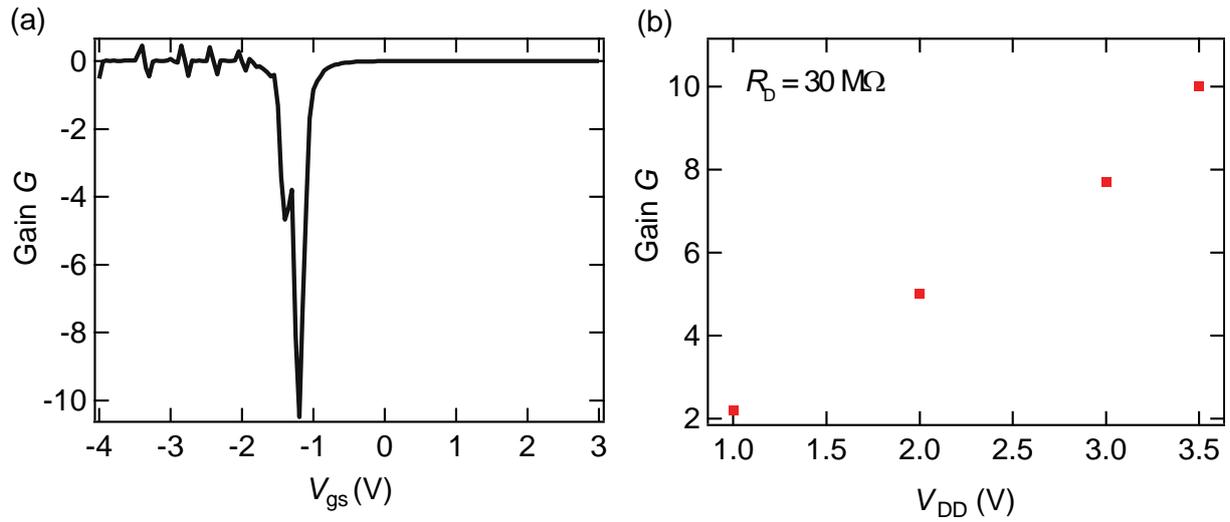

**Figure S3.** Gain of the small-signal single-transistor amplifier. (a) Gain of the amplifier from the Figure 2b measured as $G = \frac{\Delta V_{out}}{\Delta V_{in}}$. Gain is negative since we have the amplifier in common-source configuration. (b) Demonstration of the possibility of tuning the gain of the amplifier by changing the power supply $V_{DD}$.